\documentclass[
reprint,
superscriptaddress,
nofootinbib,
amsmath,amssymb,
aps,
prx,
floatfix,
showkeys
]{revtex4-2}

\usepackage{csquotes}
\usepackage{graphicx}
\usepackage{acronym}
\usepackage{bm}
\usepackage{xspace}
\usepackage[table,dvipsnames]{xcolor}
\usepackage{float}
\usepackage[caption=false]{subfig}
\usepackage[tableposition=top]{caption}
\usepackage[
  colorlinks=true,
  linkcolor=blue,
  citecolor=blue,
  urlcolor=blue,
  breaklinks=true
]{hyperref}

\newacro{ode}[ODE]{ordinary differential equation}
\newacro{eoc}[EoC]{edge of chaos}
\newacro{mips}[MIPS]{motility-induced phase separation}
\newacro{ep}[EP]{entropy production}

\newcommand{\D}{\text{d}}

\newcommand{\affilsimtech}{Stuttgart Center for Simulation Science, Cluster of Excellence EXC 2075, University of Stuttgart, Universit\"atsstra{\ss}e 32, 70569 Stuttgart, Germany}
\newcommand{\affilwin}{WIN-Kolleg of the Young Academy $\vert$ Heidelberg Academy of Sciences and Humanities, Karlstra{\ss}e 4, 69117 Heidelberg, Germany}
\newcommand{\affiltuebingen}{Institute for Theoretical Physics, University of Tübingen, Auf der Morgenstelle 14, 72076 Tübingen}
\newcommand{\affilgarching}{Max Planck Institute for Plasma Physics, Boltzmannstra{\ss}e 2, 85748 Garching, Germany}

\begin{document}

\title{
Entropy production of active matter systems as indicator for computing performance
}

\author{Patrick Egenlauf}
\email{patrick.egenlauf@simtech.uni-stuttgart.de}
\affiliation{\affilsimtech}

\author{Hannes A. Kröninger}
\affiliation{\affilsimtech}
\affiliation{\affilgarching}

\author{Arnulf Kung}
\affiliation{\affilsimtech}
\affiliation{\affiltuebingen}

\author{Mario U. Gaimann}
\affiliation{\affilsimtech}

\author{Miriam Klopotek}
\affiliation{\affilsimtech}
\affiliation{\affilwin}

\date{\today}

\begin{abstract}

Physical systems can process information through their natural dynamics, offering alternatives to conventional digital computing. Reservoir computing offers a basic framework by using a nonlinear substrate to map inputs into rich dynamical states read out by a simple linear layer. Active matter substrates are striking examples; they continuously consume energy and produce entropy. Theoretically, \ac{ep} can describe the irreversibility and distance from equilibrium. But it remains unclear whether it can track computational capabilities. We address this conceptual gap by analyzing a driven swarm reservoir model. The system \ac{ep} is computed from phase-space contraction and the bath \ac{ep} from heat flow, separately, and put in direct association to prediction performance on a Lorenz-63 task. Via force parameter scans, we show that dynamical regimes with the strongest response to a driver as well as dissipation coincide with peak performance. Therein, the dynamical discrepancy between innate (minimal dissipation) and driven transferred heat (maximal dissipation) is sharpest. Generally, driver work and relative differences of driven and undriven \ac{ep} closely mirror the performance landscape. The system \ac{ep}, derived from a generalized Liouville-equation estimator, and heat flow provide complementary diagnostics and metrics, which are most robust in the best-performing regime. These results extend prior expectations that dissipation matters for computation by identifying when and how it becomes predictive. They also relate inference power to innate dynamics, pointing to generic principles for physical computing and where \ac{ep} offers a screening metric for reservoirs and other base substrates.

\end{abstract}


\maketitle


\section{Introduction}

Physical systems can process information through their intrinsic dynamics, offering a route to computing beyond conventional digital architectures~\cite{hermans2015Trainable,jaeger2023formal}.
In this context, reservoir computing has emerged as a particularly promising framework for tasks such as chaotic time-series prediction and classification, because a rich nonlinear substrate can transform inputs into a high-dimensional dynamical representation that is read out with simple linear methods~\cite{tanaka2019recent}.
Examples of physical substrates for reservoir computing include active matter~\cite{gaimann2025robustly,wang2024Harnessing,zhou2026scalable,gaimann2026optimal,lymburnReservoirComputingSwarms2021a}, photonic systems~\cite{vandoorne2008toward,duport2012alloptical}, spintronic devices~\cite{marrows2024Neuromorphic,chen2025spintronic}, mechanical systems~\cite{coulombe2017computing,kiyabu2025optomechanical,perkins2025mechanical}, and many others~\cite{fernando2003pattern,obst2013nanoscale,nguyen2020reservoir,zhu2025Practical}.

In these cases, `inference' generally equivocates to predicting the future state of a system's environment, which drives its dynamics. But there is a need for basic principles to clarify which substrates might entail high inference power under a variety of conditions. Thus, a central question to understanding physical computing must be how computational capability relates to non-equilibrium thermodynamics~\cite{wolpertStochasticThermodynamicsComputation2019}.
As a first, computation in real physical substrates is never free: it is constrained by irreversibility, dissipation, and \acf{ep}, which quantify the extent to which a process breaks time-reversal symmetry~\cite{chattopadhyay2025Landauer,nardiniEntropyProductionField2017}.
This connection makes \ac{ep} a conceptually appealing candidate for characterizing physical information processing.

Basic links between \ac{ep} and computational performance or inference power are motivated by several related lines of work.
Since Landauer's principle, information processing has been linked to heat dissipation and irreversibility: erasing one bit requires a minimum heat dissipation of $Q = k_\mathrm{B}T\ln 2$ into the environment~\cite{landauerIrreversibilityHeatGeneration1961}.
This idea has been generalized to broader classes of computations and nonequilibrium computational models~\cite{sagawaThermodynamicsInformationProcessing2012,sagawaThermodynamicLogicalReversibilities2014,wolpertStochasticThermodynamicsComputation2019}.
These results concern thermodynamic costs rather than computational capability, but they establish \ac{ep} as a natural quantity for characterizing physical information processing.
Independently, entropy production has been shown to exhibit signatures of nonequilibrium phase transitions, including critical behavior of the entropy-production rate or its derivatives in lattice models~\cite{tomeEntropyProductionNonequilibrium2012,noaEntropyProductionTool2019,martynecEntropyProductionCriticality2020}.
At the same time, criticality has long been associated with enhanced computational capability in physical and biological systems, often discussed in terms of the \ac{eoc} between ordered and disordered dynamics~\cite{packardAdaptationEdgeChaos1987,langtonComputationEdgeChaos1990,teuscherRevisitingEdgeChaos2022}.
Similar ideas have also been explored in reservoir computing, where echo-state networks and random Boolean reservoirs can show enhanced information processing near transitions between dynamical regimes~\cite{boedeckerInformationProcessingEcho2012,snyderComputationalCapabilitiesRandom2013}.

Physical computation can also be viewed through the lens of self-organization: complex, interacting systems respond to external driving forces, and this response can be harnessed for information processing. Historically, the empirical observation that diverse systems autonomously reach states that appear `novel' or `useful' (e.g., ordered patterns, functional structures) gave rise to the idea that nature may follow an optimization principle rooted in thermodynamics~\cite{Onsager1931a,Onsager1931b}. For example, a linear non-equilibrium system in a steady state may follow a principle of minimum entropy production~\cite{Prigogine1947,Klein1954}, while dissipative systems may instead organize their flows to maximize the (local) rate of thermodynamic entropy production~\cite{Niven2009}. However, it remains unclear whether these or similar variational principles~\cite{Onsager1953,PALTRIDGE1979,Dewar2003,Dewar2005} apply to physical computing, either in practice or in theory.

Active matter models provide a test bed for exploring these connections. The dynamics are microscopically accessible yet macroscopically rich; the continuous non-equilibrium dissipation at the particle level coincides with the emergence of collective dynamics, which can be harnessed for computation.
By continuously consuming energy at the microscopic level, active matter systems reside far from equilibrium and generically produce entropy even in steady state~\cite{mandalEntropyProductionFluctuation2017,nardiniEntropyProductionField2017}.
Their \ac{ep} has therefore become an important observable for quantifying non-equilibrium behavior, with recent work demonstrating methods to measure local \ac{ep} and relate it to the underlying dynamics~\cite{roModelFreeMeasurementLocal2022}.
More generally, it has been shown that \ac{ep} and probability currents indicate how strongly a system departs from equilibrium and how that departure is distributed across the degrees of freedom~\cite{boffi2024Deep}.
While these studies establish \ac{ep} as a robust marker of non-equilibrium behavior, its relation to the information-processing ability of such systems remains largely unexplored.

In this paper, we ask whether \ac{ep} can serve not only as a thermodynamic descriptor, but also as an indicator of computing performance in driven active matter systems.
Specifically, we consider a driven active matter system and analyze its \ac{ep} in relation to its ability to support computation.
We focus on an active swarm reservoir, for which previous studies have shown that computational performance is maximized in a near-critically damped regime~\cite{gaimann2025robustly,gaimann2026optimal}.
However, such regimes are usually identified empirically through task performance.
It remains unclear whether they also carry a thermodynamic signature and, if so, which entropy channel is relevant: intrinsic phase-space contraction, heat-like dissipation, or work injected by the input drive.

We address three questions.
First, how can entropy rates be defined and estimated for a deterministic active swarm with non-Hamiltonian forces?
Second, how do system entropy rate and heat flow change between the different regimes?
Third, do these entropy rates explain reservoir performance, or do they merely reflect large energy input?

We start by reviewing recent approaches to quantifying \ac{ep} in active matter systems, and introduce our approach for calculating \ac{ep} in the context of a deterministic active matter model in Sec.~\ref{sec:theory}.
We then present our methods for simulating the system and computing \ac{ep} in Sec.~\ref{sec:methods}, followed by results on the relationship between \ac{ep} and computational performance across different dynamical regimes in Sec.~\ref{sec:results}.
Finally, we discuss the implications of our findings and potential future directions in Sec.~\ref{sec:discussion}.


\section{Thermodynamic and Computational Foundations}
\label{sec:theory}

\subsection{\Acl{ep} in active-matter systems}
Over the past years, there has been a growing interest in computing the rate of \ac{ep} in active-matter systems~\cite{fodorIrreversibilityBiasedEnsembles2022,fodorHowFarEquilibrium2016,nardiniEntropyProductionField2017,roModelFreeMeasurementLocal2022,catesStochasticHydrodynamicsComplex2022,cocconiEntropyProductionExactly2020}. The key motivation for considering the \ac{ep} rate is to quantify how far a non-equilibrium system is from an equilibrium state~\cite{fodorHowFarEquilibrium2016,nardiniEntropyProductionField2017,fodorIrreversibilityBiasedEnsembles2022}. For example, Ref.~\cite{fodorHowFarEquilibrium2016} could show that, if the persistence time of the noise governing the active behavior of the system is small, the \ac{ep} rate vanishes and thus the model is in an effective equilibrium state. As another example, the phenomenon of \ac{mips} has been analyzed via the computation of the \ac{ep} rate~\cite{nardiniEntropyProductionField2017,roModelFreeMeasurementLocal2022}. Due to the lack of large-scale mass currents typical for non-equilibrium phenomena, it was not obvious how to precisely characterize the non-equilibrium nature of \ac{mips}. However, the non-vanishing \ac{ep} rate gives an unambiguous characterization of this fact, leading also to the conclusion that mappings to equilibrium are rather limited possible.

Most works start with the following definition of the \ac{ep} rate of an active-matter system, which is based on the results of stochastic thermodynamics~\cite{seifertEntropyProductionStochastic2005a,seifertStochasticThermodynamicsFluctuation2012c},
\begin{equation}
	\label{eq:ep_active_matter}
	\sigma :=\lim_{t\rightarrow\infty}\frac{1}{t}\left\langle\ln\frac{\mathcal{P}}{\mathcal{P}^R}\right\rangle.
\end{equation}
This definition amounts to comparing the probability $ \mathcal{P} $ of a trajectory with that of its time-reversed version $ \mathcal{P}^R $ (choosing the time-reverse can in fact be subtle, cf.~\cite{fodorIrreversibilityBiasedEnsembles2022}). Defined in such a way, the \ac{ep} rate has a clear interpretation. If the occurring trajectories are more likely than their time-reversed versions, the system is irreversible, since some processes are more likely to occur in one direction of time than in the opposite. Accordingly, a non-vanishing \ac{ep} rate is primarily interpreted as signature that there are time-irreversible processes important for a system's behavior~\cite{nardiniEntropyProductionField2017,roModelFreeMeasurementLocal2022}. In principle, it is even possible to define Eq.~\eqref{eq:ep_active_matter} as some general measure of irreversibility without a clear relation to the actual physical \ac{ep} rate; when exactly such a relation exists depends on the details of the model under consideration (cf. again~\cite{fodorIrreversibilityBiasedEnsembles2022}).

Unfortunately, Eq.~\eqref{eq:ep_active_matter} is not an appropriate definition for the model we aim to work with, which is governed by a deterministic, Newtonian, but non-Hamiltonian time evolution, where the coupling to the bath and the activity of the system is modeled via non-conservative forces. The reason for Eq.~\eqref{eq:ep_active_matter} being inappropriate lies less in the fact that it has been primarily applied to stochastic models~\cite{nardiniEntropyProductionField2017,seifertEntropyProductionStochastic2005a,seifertStochasticThermodynamicsFluctuation2012c,fodorHowFarEquilibrium2016,lebowitzGallavottiCohenType1999}; indeed, there do exist corresponding results for Hamiltonian~\cite{kawaiDissipationPhaseSpacePerspective2007}, or thermostatted Nose-Hoover dynamics~\cite{monnaiMicroscopicReversibilityHeat2013}, which are both deterministic, and probabilities only arise via the dependence on an initial ensemble. Rather, the reason is that the dynamics of our system is not time-reversal invariant, in the sense that for any possible trajectory also its time-reversed version is a solution of the equations of motion. This is the case for all other dynamics where Eq.~\eqref{eq:ep_active_matter} is applied to, where the time-reversed trajectory is a (possibly) improbable, but viable solution to the equations of motion. If that is not the fact, obviously the expression~\eqref{eq:ep_active_matter} diverges since the denominator vanishes.

\subsection{\Acl{ep} in the swarm system}
We thus start our considerations on the \ac{ep} rate of the swarm system at a different point. In general, since the entropy is additive, the entropy $ S_\text{tot} $ of a composite total system that consists of the system under consideration itself and an unresolved environment, may be written as a sum of their individual entropies $ S_\text{sys} $ and $ S_\text{env} $, and, accordingly, its rate of change $ \dot{S}_\text{tot} $ consists of two parts,
\begin{equation}
	\label{eq:eprate_total}
	\dot{S}_\text{tot}=\dot{S}_\text{sys}+\dot{S}_\text{env},
\end{equation}
whose sum must be greater than, or equal to, zero, due to the second law of thermodynamics. Below, we will describe our approach to identify these two contributions.

The equations of motion of the swarm system have the general form
\begin{subequations}
	\label{eq:eom_hamiltonian}
	\begin{align}
		&\frac{\text{d}q_{i\alpha}}{\text{d}t}=\frac{\partial}{\partial p_{i\alpha}}H(Q,P,t),\\
		&\frac{\text{d}p_{i\alpha}}{\text{d}t}=-\frac{\partial}{\partial q_{i\alpha}}H(Q,P,t)+B_{i\alpha}(Q,P,t),
	\end{align}
\end{subequations}
where $ q_{i\alpha} $, $ p_{i\alpha} $ are the $ \alpha $-th component of the generalized position (momentum resp.) coordinate of the $ i $-th particle, $ Q $, $ P $ denote the set of all position and momentum coordinates. $ H(Q,P,t) $ is a time-dependent Hamiltonian, which includes all forces that can be written as the gradient of a potential, and $ B_{i\alpha}(Q,P,t) $ are functions accounting for all non-conservative forces in the system. Let further denote $ \Omega=(Q,P) $.

The system's internal entropy (production) may be regarded simply as the Gibbs entropy
\begin{equation}
	\label{eq:gibbsentropy_def}
	S_\text{sys}(t):=-\int\rho_t(\Omega)\log\rho_t(\Omega)d\Omega,
\end{equation}
where the system's state is described by a time-dependent phase-space density $ \rho_t(\Omega) $. Since the dynamics is not purely Hamiltonian, Liouville's theorem preventing the Gibbs entropy to change does not hold; in fact, the presence of non-conservative forces lead to a change of it. However, one should be cautious to identify it with the actual thermodynamic entropy of the system. This is for one reason because of the foundational issues concerning the Gibbs entropy ~\cite{goldsteinGibbsBoltzmannEntropy2020a}, for the other because the swarm model is not a thermodynamically complete model, since the coupling to the bath does not respect the thermodynamic properties of the bath (unlike, e.g., in the case of a Langevin dynamics, where the diffusion constant is related to the bath's temperature).

Nevertheless, it provides useful insights into the physics of the system. This is because the time derivative $ \dot{S}_\text{sys} $ of~\eqref{eq:gibbsentropy_def} gives the rate of phase-space contraction, i.e., how much the volume occupied by a phase-space density de- or increases over time. In other words, it quantifies how distinct initial conditions are separated further away or compressed together under time evolution. Accordingly, there is a relation ~\cite{daemsEntropyProductionPhase1999} to the system's Lyapunov exponents $ \lambda_i $,
\begin{equation}
	\lim_{t\rightarrow\infty}\left\langle \dot{S}_\text{sys}\right\rangle_t=\sum_{i}\lambda_i,
\end{equation}
which means that the time average over the EP rate converges to the sum of all Lyapunov exponents of the system.

The second component in Eq.~\eqref{eq:eprate_total} should be considered as consisting of two parts. This is because the constituent particles of active-matter systems are not merely passive, but are itself small subsystems like, e.g., bacteria or birds. Inside them, there may happen rather complex processes that might produce a significant amount of entropy, but which are unresolved in usual active-matter models. Hence, additional to a bath of a liquid or a gas the system is suspended in, these internal degrees of freedom of the constituents should be considered as being part of the unresolved environment. Its \ac{ep} rate thus reads
\begin{equation}
	\label{eq:eprate_env_splitting}
	\dot{S}_\text{env}=\dot{S}_\text{bath}+\dot{S}_\text{cons}.
\end{equation}
If one aimed at computing the second contribution, one would have to introduce a physical model of the constituents' internal processes. Since there might exist more than one suitable model, this would lead to some degree of arbitrariness in the computation of the \ac{ep} rate. We would like to avoid this in order to focus on the other contributions, which are more unambiguously determined and perhaps more interesting for our purposes. However, we will keep this point in mind for the discussion of our results.

The other part of the environment's \ac{ep} rate amounts to a change of entropy in the heat bath. It is reasonable to assume that the bath is composed in such a way that it changes its entropy only due to interactions with the system, and that its internal processes occur on a much smaller timescale than that of the system itself. Hence, one may assume that the change in entropy is given by the flow of heat from the system to the bath,
\begin{equation}
	\label{eq:ep_flow_general}
	\dot{S}_\text{bath}=\frac{\dot{Q}}{T}.
\end{equation}
We propose the following, physically motivated identification of the heat: Taking the time derivative of the Hamiltonian, aka the total energy of the system, yields
\begin{equation}
	\label{eq:total_energy_derivative}
	\frac{\text{d}E}{\text{d}t}=\frac{\text{d}H}{\text{d}t}=\frac{\partial H}{\partial t}+\dot{q}_{i\alpha}B_{i\alpha}.
\end{equation}
The first part corresponds to a change of energy due to an interaction with an external driving. The second part corresponds to the energy change due to the non-conservative forces, which model both, the active behavior of the system's constituents, and the interaction with the bath. Only the latter part leads to an entropy increase in the bath. In order to separate the two contributions, we propose the following approach. Suppose that $ q_{i\alpha} $ corresponds to Cartesian coordinates, such that $ \dot{\bm{q}}_i=(\dot{q}_{i,1},\dot{q}_{i,2},...) $ is the Cartesian velocity vector of the particle $ i $. Hence, if the vector $ \bm{B}_i=(B_{i,1},B_{i,2},...) $ points opposite to $ \dot{\bm{q}}_i $, particle $ i $ is decelerated by the non-Hamiltonian forces, and thus energy taken from the system and put into the bath. Thus, we assume that only these negative parts contribute to a heat flow, leading to the definition
\begin{equation}
	\label{eq:heat_by_eom}
	\dot{Q}=-\sum_{i,\alpha}\dot{q}_{i\alpha}B_{i\alpha}\Theta\left(-\dot{\bm{q}}_{i}\cdot\bm{B}_{i}\right).
\end{equation}
Note that this choice corresponds to a purely `passive' bath. It can only absorb heat from the system, and not itself couple back to it, unlike in the case of, e.g., Langevin dynamics where the coupling to the bath is realized by a Gaussian white noise which may take either positive or negative sign. All other contributions to the non-conservative forces are ascribed to the active constituents, whose internal dynamics is decoupled from the bath. By definition, the expression~\eqref{eq:heat_by_eom} and hence \eqref{eq:ep_flow_general} is positive, i.e., there is always a net flow of heat from the system to the bath.

Given these definitions, it would be now an interesting question if these were consistent with the second law of thermodynamics, i.e., if Eq.~\eqref{eq:eprate_total} with those definitions was positive. Unfortunately, it is not clear if this is the case. The problem is that it is unclear whether the swarm model is thermodynamically consistent. In order to prove that, it would be necessary to employ a model which relates the thermodynamic properties of the bath with the interactions with the system, like it is the case, for example, for Langevin dynamics, where the diffusion coefficient is related to the bath's temperature by the Einstein relations~\cite{seifertStochasticThermodynamicsFluctuation2012c}. One way to account for this fact would be to introduce an artificial model for these couplings, as has been done for Nose-Hoover dynamics~\cite{espositoNonequilibriumThermodynamicsNoseHoover2011}. However, even if this would be possible in some natural way for the swarm system, there would still be the problem that it is not clear how the constituents' internal degrees of freedom contribute to the total \ac{ep} rate. For these reasons, we will not consider this point further and restrict ourselves considering $ \dot{S}_{sys} $ and $ \dot{S}_{bath} $ separately, since these two quantities already contain much information on the physics of our system.


\section{Swarm Dynamics and its Entropy-Rate Calculation}
\label{sec:methods}

\subsection{Swarm System}
Swarm systems were first employed for reservoir computing by~\cite{lymburnReservoirComputingSwarms2021a}, whose proposed equations of motions we will use. A system of $ n $ particles is subject to the interaction forces
\begin{subequations}

\begin{eqnarray}
	&\bm{F}^\text{r}_k&=\sum_{j\neq k}\Theta(r_\text{r}-\|\bm{x}_k-\bm{x}_j\|)\frac{\bm{x}_k-\bm{x}_j}{\|\bm{x}_k-\bm{x}_j\|^2},\\
	\label{eq:alignment_force_def}
	&\bm{F}^\text{a}_k&=\sum_{j\neq k}\Theta(r_\text{a}-\|\bm{x}_k-\bm{x}_j\|)(\dot{\bm{x}}_j-\dot{\bm{x}}_k),\\
	&\bm{F}^\text{h}_k&=\bm{x}^\text{h}_k-\bm{x}_k,\\
	\label{eq:speed-controller_force_def}
	&\bm{F}^\text{sc}_k&=-\dot{\bm{x}_k}\frac{(\|\dot{\bm{x}}_k\|-s)}{s},
\end{eqnarray}
\end{subequations}
where $ \bm{F}^\text{r} $ is a Coulomb-like repulsive interaction, which is only active in a certain local radius $ r_\text{r} $, denoted by the $ \Theta $ function (with the convention that $ \Theta(x)=1 $ for $ x\geq0 $ and $ \Theta(x)=0 $ otherwise); $ \bm{F}^\text{a} $ an alignment force between the velocities of particles within a radius $ r_\text{a} $; $ \bm{F}^\text{h} $ a `homing' force attracting each particle to its homing position $ \bm{x}^\text{h}_k $; and $ \bm{F}^\text{sc} $ a speed-controller force aiming to keep the magnitude of each particle's velocity to a fixed value $ s $.

The signal is coupled into the system by a distinguished particle, a `driver', whose trajectory $ \bm{x}_\text{d}(t) $ is determined externally. There is a repulsive force between the driver and the other particles within a radius of $ r_\text{d} $,
\begin{equation}
	\label{eq:driver_force}
	\bm{F}^\text{d}_k=\Theta(r_\text{d}-\|\bm{x}_k-\bm{x}_\text{d}\|)\frac{\bm{x}_k-\bm{x}_\text{d}}{\|\bm{x}_k-\bm{x}_\text{d}\|^2}.
\end{equation}
Thus, the total force exerted on particle $ k $ reads
\begin{equation}
	\label{eq:forces_sum}
	\bm{F}_k=K_\text{r}\bm{F}^\text{r}_k+K_\text{a}\bm{F}^\text{a}_k+K_\text{h}\bm{F}^\text{h}_k+K_\text{sc}\bm{F}^\text{sc}_k+K_\text{d}\bm{F}^\text{d}_k,
\end{equation}
where $ K_\text{r, a, h, sc, d} $ are coupling constants. The system evolves according to the Newtonian equation of motions
\begin{equation}
	\label{eq:eom_general}
	m_k\ddot{\bm{x}}_k=\bm{F}_k,
\end{equation}
where the masses were set to unity by~\cite{lymburnReservoirComputingSwarms2021a}, but will be kept here for the sake of generality. According to the original suggestion of~\cite{lymburnReservoirComputingSwarms2021a}, it is possible to introduce an additional $ \tanh $-wrapper
\begin{equation}
	\label{eq:tanh_wrapper_def}
	\bm{F}_k^{w}=\alpha\tanh\left(\beta\bm{F}_k\right)
\end{equation}
to keep the dynamics bounded. In the context of \ac{ep}, this poses a problem concerning the identification of the entropy flow into the environment, as will be discussed in Sec.~\ref{sec:methods:ep:ep_env}. Therefore, instead of the $ \tanh $-wrapper we use a force capper that is linear for small forces, but keeps the dynamics bounded by capping the force at a certain value.
With a cap value $ F_\mathrm{cap}>0 $, we define
\begin{equation}
	\label{eq:force_capper_def}
	\tilde{\bm{F}}_k=\min\left\{1,\frac{F_\mathrm{cap}}{\|\bm{F}_k\|}\right\}\bm{F}_k.
\end{equation}
Hence, for $ \|\bm{F}_k\|\leq F_\mathrm{cap} $ the dynamics is unchanged, while for $ \|\bm{F}_k\|>F_\mathrm{cap} $ the force magnitude is clipped to $ F_\mathrm{cap} $. This keeps the equations bounded and only minimally affects the calculation of the \ac{ep}, since the forces are only capped in extreme cases. These are rare, which we show in Appendix~\ref{app:force-capper}, and thus do not contribute much to the overall \ac{ep}.

The equations may be rewritten in the Hamiltonian form~\eqref{eq:eom_hamiltonian}, with time-dependent Hamiltonian
\begin{equation}
	\label{eq:hamiltonian_swarm}
	H(Q,P,t)=\sum_{i=1}^{n}\frac{p_i^2}{2m_i}+V(Q)+V_{\text{ext}}(Q,t),
\end{equation}
where
\begin{subequations}
	\label{eq:potential_def}
	\begin{align}
		&V(Q)=\\
		&\frac{1}{2}\sum_{i\neq j}\left(\frac{K_\text{r}}{\|\bm{q}_i-\bm{q}_j\|}+c_\text{r}\right)\Theta(r_\text{r}-\|\bm{q}_i-\bm{q}_j\|)\\
		&-\sum_i\frac{1}{2}K_\text{h}(\bm{x}_{\text{h}_{i}}-\bm{q}_{i})^2
	\end{align}
\end{subequations}
is the inter-particle interaction potential including homing and repulsion force, and
\begin{equation}
	\label{eq:potential_ext_def}
	V_{\text{ext}}(Q,t)=\sum_i\left(\frac{K_\text{d}}{\|\bm{q}_i-\bm{x}_{\text{d}}(t)\|}+c_\text{d}\right)\Theta(r_\text{d}-\|\bm{q}_i-\bm{q}_{\text{d}}\|),
\end{equation}
the interaction potential with the driver. By choosing appropriate constants $ c_\text{r} $, $ c_\text{d} $, both potentials become continuous. The two forces which cannot be expressed via a Hamiltonian are the alignment and the speed-controller force; the functions
\begin{subequations}
	\begin{align}
		&B_{i\alpha}(Q,P,t)\\
		&=K_\text{a}\sum_{j\neq i}\Theta\left(r_\text{a}-\|\bm{q}_i-\bm{q}_j\|\right)\left(\frac{p_{j\alpha}}{m_j}-\frac{p_{i\alpha}}{m_i}\right)\\
		&+K_\text{sc}p_{i\alpha}\frac{\|\bm{p_i}\|-m_i s}{m_i^2 s}
	\end{align}
\end{subequations}
account for these forces when being included into the equations of motions as in ~\eqref{eq:eom_hamiltonian}. As usual, the Hamiltonian $ H(Q,P,t) $ gives the total energy as sum of the kinetic and potential energies.

\subsection{Reservoir Computing}
Reservoir Computing is a paradigm for performing computations in real time, and has been first proposed for analyzing information processing in biological systems~\cite{maassRealTimeComputingStable2002a,jaegerHarnessingNonlinearityPredicting2004a}. However, since then the idea has been extended to other dynamical systems acting as a substrate for computation, of which some have even been implemented experimentally~\cite{wang2024Harnessing,duport2012alloptical,marrows2024Neuromorphic,kiyabu2025optomechanical,perkins2025mechanical,fernando2003pattern,obst2013nanoscale,nguyen2020reservoir}.

The general idea is to couple a temporally continuous input $ u(t) $ into a dynamical system $ x(t) $ of much larger dimension than the input signal. By responding to it, the system creates states in its high-dimensional state space, which thus contain information the whole history of inputs. In order to make use of the information, one reads out the system's state by some observational variables $ r(t)=r(x(t)) $, which are functions of the full state. Mathematically, this can be described as the input sequence being mapped to the observed sequence by a filter, $ r(t)=F[u(t)] $. Finally, a linear map $ L_w $ is applied to these observations, such that
\begin{equation}
	y(t)=L_wr(t),
\end{equation}
where $ y(t) $ is the desired output signal which was to be computed. The crucial point is that training the model is only required in this last point, using a simple linear regression, whereas the dynamical system follows its free, untrained dynamics. This property has made reservoir computing a promising ansatz to approach hard-to-train problems concerning continuous-time input sequences~\cite{tanaka2019recent,schrauwen2007overview,cucchi2022hands}.

Since in principle any system can serve as a substrate for reservoir computing, there have been many suggestions to use various physical systems as reservoir~\cite{coulombe2017computing,zhu2025Practical,vandoorne2008toward,chen2025spintronic}. Swarm systems have first been introduced as reservoirs by~\cite{lymburnReservoirComputingSwarms2021a}. In order to do so, it is necessary to (i) decide how to couple the input signal with the reservoir, and (ii) how to read out the system's state to further apply the linear map on it.
\begin{figure}
    \centering
    \includegraphics[width=\columnwidth]{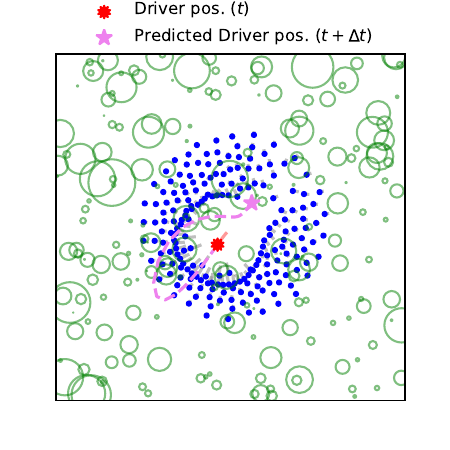}
    \caption{Representative snapshot of the swarm reservoir setup. Blue dots are agents ($n = 200$), the red dot is the predator (input) and green circles show Gaussian readout kernels (centers $\mathbf{c}_m$ and widths $\sigma_m$). The linear readout is trained to predict the predator position at $ t+\Delta t $ (pink star).}
    \label{fig:swarm_reservoir_example}
\end{figure}

Concerning (i),~\cite{lymburnReservoirComputingSwarms2021a} introduced the driver into the swarm, coupled to the agents via Eq.~\eqref{eq:driver_force}. Externally prescribing the driver's position thus provides the possibility to inject a $ d $-dimensional (in~\cite{lymburnReservoirComputingSwarms2021a} in fact $ d=2 $) signal via $ x_\text{d}(t)=u(t) $ into the system.

Concerning (ii), in principle one could use directly the full $ 2dN $-dimensional point in the system's phase space with all microscopic positions and momenta of the particles as observables. However, as~\cite{lymburnReservoirComputingSwarms2021a} pointed out, this does not work very well. They suspected that the reason was the permutation symmetry of the system, since all particles have the same mass. Instead, the introduction of observational kernels has been proposed. One takes a number $ M $ of Gaussian functions
\begin{equation}
	\psi_m=\exp\left(\frac{(\mathbf{x}-\mathbf{c}_m)^2}{2\sigma^2_m}\right),
\end{equation}
centered at some positions $ \mathbf{c}_m $ in the $ d $-dimensional position space, and having some widths $ \sigma_m $, with $ m=1,...,M $. Then one can define the $ \psi_m $-weighted average number
\begin{equation}
	r^{(n)}_m(t)=\sum_{k}\psi_m(\mathbf{x}_k(t)),
\end{equation}
and velocity
\begin{equation}
	r^{(v_i)}_m(t)=\sum_{k}\psi_m(\mathbf{x}_k(t))\dot{x}_{k,i}(t).
\end{equation}
An example of such a setup is shown in Fig.~\ref{fig:swarm_reservoir_example}. The linear layer is then applied to these $ 2dM $-dimensional observations instead of the full phase space.
This prevents the problems involved with the permutation symmetry.

The swarm system shows interesting different dynamical regimes, which may be connected with the computational properties; for further discussions see~\cite{lymburnReservoirComputingSwarms2021a} and~\cite{gaimann2025robustly}.

\subsection{\Acl{ep}}

In order to numerically evaluate the two contributions $ \dot{S}_\text{sys} $ and $ \dot{S}_\text{bath} $ to the total rate of entropy production~\eqref{eq:eprate_total} we employ two different methods. For the system's internal entropy, we suggest a Monte-Carlo approach based on the generalized Liouville equation, which governs the time evolution of the density $\rho_t$. This method can be applied to arbitrary systems of differential equations, and thus seems to be interesting in its own. The entropy flow into the environment is determined via an analysis of the microscopic equations of motion, and the identification of terms that contribute to a heat transfer.

\subsubsection{Internal \ac{ep}: The Generalized Liouville Equation approach}

Numerically computing Eq.~\eqref{eq:gibbsentropy_def} for a phase space of large dimension can, in general, be a quite demanding task. Either one has to solve a partial differential equation governing the evolution of the density, or one has to estimate $ \rho $ empirically from a large number of samples~\cite{verduEmpiricalEstimationInformation2019,valiantEstimatingUnseenImproved2017}, becoming again infeasible for high dimensions.

Therefore, we propose an estimator which is based on the fact that the phase space density evolves according to a first-order transport equation. This allows us to rewrite Eq.~\eqref{eq:gibbsentropy_def} in such a way that it is not necessary to solve the full high-dimensional partial differential equation, but to evaluate the density along a number of independent trajectories and then compute Eq.~\eqref{eq:gibbsentropy_def} and its derivative using the Monte-Carlo method. Since this method is valid for any system of first-order \acp{ode},
\begin{equation}
	\label{eq:eom_dynamicalsystem_general}
	\frac{d\xi_i}{dt}=F_i(\xi,t),
\end{equation}
with $ \xi=(\xi_1,...,\xi_s) $, we first derive it for this general case and later apply it to our system~\eqref{eq:eom_hamiltonian}.

The phase-space density $ \rho_t(\xi) $ evolves according to the generalized Liouville equation ~\cite{gerlichVerallgemeinerteLiouvilleGleichung1973a,steebGeneralizedLiouvilleEquation1979}
\begin{equation}
	\frac{\partial \rho_t}{\partial t}+\sum_{i=1}^{s}F_i\frac{\partial \rho_t}{\partial \xi_i}+\sum_{i=1}^{s}\rho_t\frac{\partial F_i}{\partial \xi_i}=0,
\end{equation}
which is a first-order transport equation. This allows one to write the general solution for the density $ \rho_t $ along a trajectory $ \xi(t) $ that starts at some point $ \xi_0 $ as
\begin{equation}
	\label{eq:gen_solution_rho_t}
	\rho_t(\xi(t))=\rho_0(\xi_0)\exp\left[-\int_{0}^{t}\sum_{i=1}^{s}\dfrac{\partial F_i}{\partial \xi_i}(\xi(\tau),\tau)d\tau\right]
\end{equation}
Since the volume element $ d\xi $ transforms as
\begin{equation}
	d\xi=d\xi_0\exp\left[\int_{0}^{t}\sum_{i=1}^{s}\dfrac{\partial F_i}{\partial \xi_i}(\xi(\tau),\tau)d\tau\right],
\end{equation}
it is possible to rewrite the Gibbs entropy in terms of the initial density $\rho_0$
\begin{subequations}
	\begin{align}
		S(t)=&-\int\log\left[\rho_t(\xi)\right]\rho_t(\xi)d\xi\\
		&-\int\log\left[\rho_t(\xi(t))\right]\rho_0(\xi_0)d\xi_0,
	\end{align}
\end{subequations}
with $\xi(t)$ denoting the time-evolved initial point $\xi_0$. Taking the derivative and substituting~\eqref{eq:gen_solution_rho_t} finally yields
\begin{equation}
	\label{eq:eprate_gen_initial_points}
	\dot{S}(t)=\int\sum_{i=1}^{s}\frac{\partial F_i}{\partial \xi_i}(\xi(t),t)\rho_0(\xi_0)d\xi_0.
\end{equation}

Since the function in the integrand depends only on $ t $ and $\xi(t)$, which can be simply obtained by integrating~\eqref{eq:eom_dynamicalsystem_general}, it can be evaluated point-wise, and thus allows for a Monte-Carlo estimation of the integral. In order to do so, take a number of $ N $ initial states $ \xi^{(1)},...,\xi^{(N)} $ and compute their trajectories $ \xi^{(1)}(t),...,\xi^{(N)}(t) $. The empirical Monte-Carlo estimator for~\eqref{eq:eprate_gen_initial_points} then is
\begin{equation}
	\hat{\dot{S}}(t)=\sum_{k=1}^{N}\sum_{i=1}^{s}\frac{\partial F_i}{\partial \xi_i}(\xi^{(k)}(t),t)
\end{equation}

This allows to make full use of the strengths of the Monte-Carlo integration method. The errors grow as $ \mathcal{O}(N^{-1/2}) $, independent of the dimension $ s $ of the phase space. Compare this with the other options to compute~\eqref{eq:gibbsentropy_def}. Computing the full time evolution of $ \rho_t $ numerically would require to lay a grid over the phase space. However, keeping a fixed distance, the number of grid points grows exponentially with the dimension. Alternatively, using empirical estimators for $ \rho_t $ from a large number of independent runs to compute $ \rho_t $, would require dividing the phase space $ \Gamma $ into bins. Keeping their size fixed, this requires an exponentially growing number of bins $ N_{bin} $. Since on the other hand all empirical estimator for the entropy require a number of $ \mathcal{O}(N_{bin}/\log N_{bin}) $ samples~\cite{verduEmpiricalEstimationInformation2019}, their number increases exponentially too, rendering this method impossible too for high-dimensional phase spaces.

For our system following the equations of motion~\eqref{eq:eom_hamiltonian}, this gives
\begin{equation}
	\label{eq:ep_sys_nonHamitonian}
	\dot{S}_{\text{sys}}(t)=\int\sum_{i,\alpha}\frac{\partial B_{i\alpha}}{\partial p_{i\alpha}}(\Omega(t),t)\rho_0(\Omega_0)d\Omega_0,
\end{equation}
where $\Omega(t)=(Q(t),P(t))$ is the trajectory in the full $ n $-particle phase space starting at $ \Omega_0 $. Hence, only the non-Hamiltonian parts of the forces, which are the alignment and the speed-controller force, contribute to the \ac{ep}, as is a well-known fact. Calculating $ \dot{S}_{\text{sys}} $ with the explicit form of the non-Hamiltonian forces gives
\begin{subequations}
	\begin{align}
		&\dot{S}_{\text{sys}}(t)=-\int\sum_{i=1}^{n}\left[D\frac{K_\text{a}}{m_i}N_i^{\text{a}}\right]\rho_0(\Omega_0)\D\Omega_0\\
		&-\int\sum_{i=1}^{n}\left[\frac{K_\text{sc}}{m_i^2s}\left((D+1)\|p_i\|-Dm_is\right)\right]\rho_0(\Omega_0)\D\Omega_0
	\end{align}

\end{subequations}
The first part is the contribution due to the alignment force, where $ N_i^{\text{a}} $ is the number of particles in the alignment neighborhood of the $ i $-th particle. Interestingly, this contribution does not depend on the values of the momenta, but only on this number of neighbors. This might be interpreted in this way that the larger the number $ N_i^{\text{a}} $ is, the more coherent become the velocities in a local neighborhood of particle $ i $, since a larger number of particles participates in the alignment, thereby lowering the entropy.

The second part is the contribution due to the speed-controller force. It may be interpreted in such a way that the contribution of particle $ i $ to the \ac{ep} rate depends on the absolute value of its velocity. If the velocity is large enough, $ v_i\geq D/(D+1)s $, the particle causes a decrease in entropy, otherwise an increase. This is because if, as in the former case, the particle is decelerated, different phase points in the $ n $-particle phase space which differ only in the velocity of particle $ i $ are moved closer together, leading to a compression of phase-space volume; vice-versa for acceleration.

In sum, the system's entropy change is caused by the non-Hamiltonian alignment and speed-controller forces, which have the two contrary effects of homogenizing the velocities by aligning and decelerating, thus decreasing the entropy, and accelerating, leading to a larger dispersion of velocities and thus increased entropy.

\subsubsection{Entropy flow into the environment}
\label{sec:methods:ep:ep_env}

Starting with the rate of change of the total energy~\eqref{eq:total_energy_derivative} and using the Hamiltonian~\eqref{eq:hamiltonian_swarm}, one obtains
\begin{equation}
	\frac{\partial H}{\partial t}=\frac{\partial V_{\text{ext}}}{\partial t},
    \label{eq:driver-work}
\end{equation}
which can be clearly identified as the work performed by the driver on the system. Using the expression~\eqref{eq:heat_by_eom} for the heat flow into the bath then gives in the ensemble average
\begin{equation}
	\dot{Q}=-\int\sum_{i=1}^{n}\frac{\bm{p}_i\cdot\bm{B}_i}{m_i}\Theta\left(-\bm{p}_i\cdot\bm{B}_i\right)\rho_0(\Omega_0)d\Omega_0,
\end{equation}
where
\begin{equation}
	\bm{p}_i\cdot\bm{B}_i=K_\text{a}\bm{p}_i\cdot\bm{\Delta V}_{i}^{\text{a}}-K_\text{sc}\bm{p}_i^2\frac{\left(\|\bm{p}_i\|-m_is\right)}{m_i^2s}.
\end{equation}
The first contribution is due to the alignment force, where
\begin{equation}
	\bm{\Delta V}_{i}^{\text{a}}=\sum_{j\neq i}\Theta\left(r_\text{a}-|q_j-q_i|\right)\left(\frac{p_j}{m_j}-\frac{p_i}{m_i}\right)
\end{equation}
is the summed difference in velocities between particle $ i $ and its neighbors. Hence, only if this difference points into the opposite direction of $ i $'s velocity, it gives a negative contribution to the scalar product. The second contribution is due to the speed-controller force, which is negative only if $ i $'s velocity is larger than the target velocity $ s $. The sum of both determines whether there is a heat flow into the bath.

Notice further that the identification of the heat entropy flow into the environment depends crucially on the distinction made between conservative and non-conservative, as they occur in the change of total energy~\eqref{eq:total_energy_derivative}. This means that this method is restricted to more specific cases than that for the computation of the system's internal \ac{ep} presented in the previous section, which is valid for any system of \acp{ode}, where it might not be possible or physically plausible to identify a coupling with an environment, or even a total energy. Accordingly, in order to compute the heat flow, it is necessary to dismiss the $\tanh$-wrapper in Eq.~\eqref{eq:tanh_wrapper_def}, because if one would apply it to the total force, it would not be possible anymore to write the equations of motion as~\eqref{eq:eom_hamiltonian}, with a clear distinction between Hamiltonian and non-Hamiltonian parts. However, using the more general version of the internal entropy production derived before, it would still be possible to determine $ \dot{S}_{\text{sys}} $, though it could not be written anymore as in~\eqref{eq:ep_sys_nonHamitonian}.


\graphicspath{{figs/}}

\section{Entropy and Performance Across Dynamical Regimes}
\label{sec:results}

We simulate $5\times 10^{5}$ time steps with a time step size of $\text{d}t=0.002$, integrated using the velocity Verlet scheme, discard the first $10^{4}$ steps as burn-in, and record forces (and thus \ac{ep} rates) every 10 steps, yielding an effective time step size of $\text{d}t_\mathrm{eff}=0.02$. Unless stated otherwise, averages and variability are computed over 25 seeds per parameter configuration for the undriven scan and 5 seeds for the driven scans, where the additional reservoir training and evaluation requires more storage. The smaller time step compared to previous work~\cite{gaimann2025robustly,gaimann2026optimal} only minimally affects the dynamics but enables a more accurate \ac{ep} estimate; therefore, absolute values are not directly comparable to prior studies, while trends remain consistent.
To match the linear regime of the $\tanh$ wrapper used in previous studies ($\alpha=200$, $\beta=0.1$), we scale the total force by 20, which rescales the $K_i$ in Eq.~\eqref{eq:forces_sum} and slightly shifts the $K_\text{sc}$ axis of the speed-controller scan relative to earlier work.
These and all other simulation parameters (e.g., box size $l_\text{box}$ and number of agents $N$) are given in Tab.~\ref{tab:swarm_parameters}. For the speed-controller scan, we vary $K_\text{sc} \in [2 \cdot 10^{-4}, 2 \cdot 10^{3}]$ and $s \in [10^{-5}, 10^{2}]$, while for the driver-force scan, we vary $K_\text{d} \in [2 \cdot 10^{-1}, 2 \cdot 10^{7}]$ and $r_\text{d} \in [0.1, 8.0]$.
\begin{table}[t]
    \caption{Parameters used in the swarm simulations. The given parameters for $K_\text{d}$, $r_\text{d}$ and $K_\text{sc}$, $s$ are used the speed-controller and driver-force scans, respectively.}
    \centering
    \begin{tabular}{c|c|c|c|c|c|c|c}
        $N$ & $M$ & $l_\text{box}$ & $l_\text{box}^\text{driver}$ & $K_\text{h}$ & $K_\text{r}$ & $K_\text{a}$ & $K_\text{sc}$ \\ \hline
        200 & 200 & 16.0 & 8.0 & 40.0 & 40.0 & 0.2 & 0.41383 \\
    \end{tabular} \\[1em]
    \begin{tabular}{c|c|c|c|c|c|c}
        $K_\text{d}$ & $r_\text{r}$ & $r_\text{a}$ & $s$ & $r_\text{d}$ & $F_\text{cap}$ & $\text{d}t$ \\ \hline
        2000.0 & 1.0 & 1.0 & 0.04833 & 2.0 & 800.0 & 0.002 \\
    \end{tabular}
    \label{tab:swarm_parameters}
\end{table}

Across the speed-controller scan ($K_\text{sc}$ versus $s$ in Eq.~\eqref{eq:speed-controller_force_def}), we distinguish three dynamical regimes.
The underdamped regime lies in the top-left corner (high target speed $s$, low speed-controller strength $K_\text{sc}$), where particles oscillate because friction is weak.
The near-critically damped regime forms a diagonal strip from bottom left (small $s$, small $K_\text{sc}$) to top right (high $s$, large $K_\text{sc}$), excluding a small region of very large $s$ and $K_\text{sc}$ where unphysical behavior occurs; here oscillations vanish and, in the undriven case, the system reaches equilibrium more quickly.
The overdamped regime occupies the bottom-right corner (low $s$, high $K_\text{sc}$), where particles move slowly along near-straight trajectories and no longer exhibit swarm-like behavior; because it interacts only weakly with the driver, it is less central to our analysis.
For more details on these regimes, see~\cite{gaimann2025robustly}.

The force capper in Eq.~\eqref{eq:force_capper_def} keeps the dynamics bounded, with $F_\mathrm{cap}=800$.
Once it is applied to a particle, the resulting force on that particle is no longer cleanly separable into Hamiltonian and non-Hamiltonian contributions, which the \ac{ep} calculation relies on.
The reduced time step helps prevent close encounters that would otherwise trigger the capper.
Before analyzing entropy rates, we verified that the force capper does not significantly influence the observed dynamics and hence the calculated \ac{ep} rates in most of the parameter space, but it can have a significant influence in the overdamped regime, which should be taken into account when interpreting the results in this regime.
For details, see Appendix~\ref{app:force-capper}.

\subsection{Time series}

\begin{figure*}[p]
    \includegraphics[width=\textwidth]{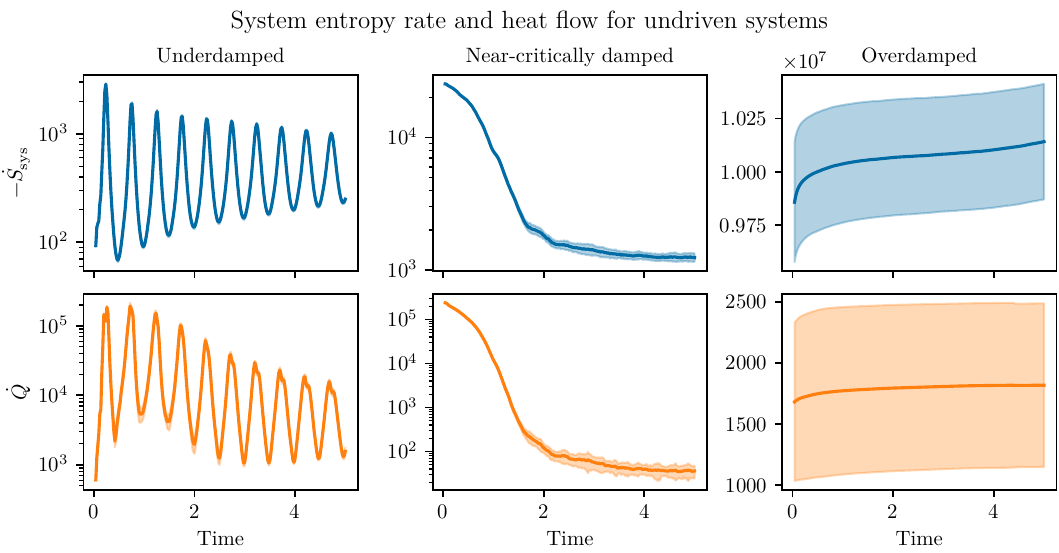}
    \includegraphics[width=\textwidth]{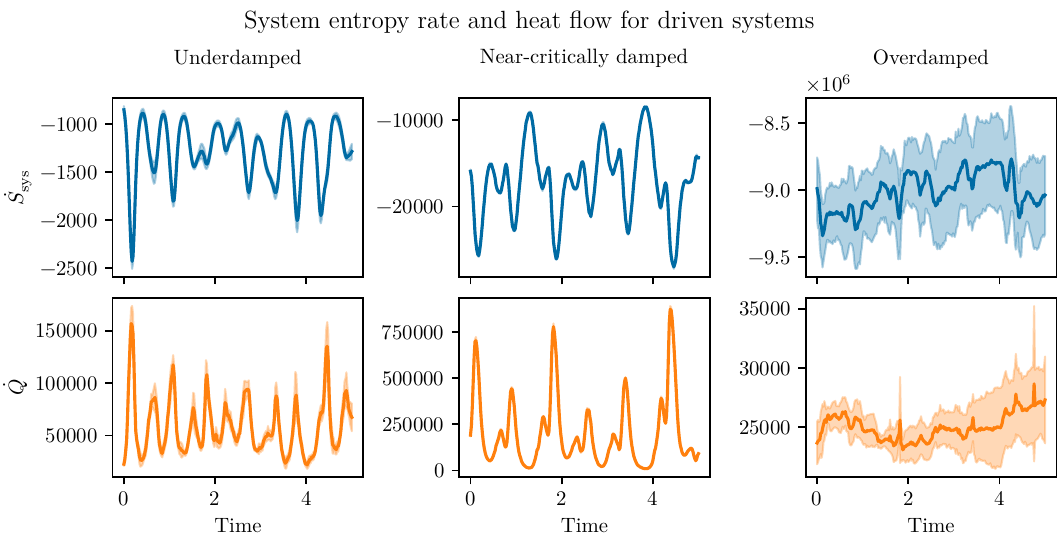}
    \caption{Time series of the system entropy rate and heat flow for the undriven and driven system for different regimes (see Tab.~\ref{tab:parameter_configurations_regimes}). For the undriven system, the first $10^4$ steps are not discarded as burn-in to show the relaxation dynamics. Only the first two steps are removed to avoid the initial condition biasing the plot. The solid lines represent the mean \ac{ep} rate over the different seeds, hence different initial conditions, and the shaded regions represent the standard deviation. The system entropy is negative for all three regimes. The difference in the dynamics of the three regimes is clearly visible in the system entropy rate and the heat flow.}
    \label{fig:entropy_time_series_driven}
\end{figure*}

We first compare time series of the system entropy rate and the heat flow in the three different regimes in the undriven and driven case, which are shown in Fig.~\ref{fig:entropy_time_series_driven}.
The respective parameter configurations used to represent the three regimes are displayed in Tab.~\ref{tab:parameter_configurations_regimes}.
\begin{table}
    \caption{Parameter configurations for the three regimes presented in Fig.~\ref{fig:entropy_time_series_driven}.}
    \centering
    \begin{tabular}{c|c|c}
        Regime & $K_\text{sc}$ & $s$ \\
        \hline
        Underdamped & 0.0011 & 18.3298 \\
        Near-critically damped & 0.4138 & 0.0483 \\
        Overdamped & 28.7690 & 0.0007
    \end{tabular}
    \label{tab:parameter_configurations_regimes}
\end{table}
In the undriven case, the system entropy rate is negative for all three regimes, but it is largest in the overdamped regime and smallest in the underdamped regime. The heat flow is positive (by definition) for all three regimes, but it is largest in the underdamped regime and smallest in the near-critically damped regime. Both the system entropy rate and heat flow show similar dynamics in the respective regimes, with the underdamped regime showing a damped oscillatory behavior, the near-critically damped regime showing a more monotonic behavior, converging towards a plateau, and the overdamped regime showing a rather constant \ac{ep} rate with comparatively large standard deviation. This could be due to a strong dependence on the initial conditions and the activity of the force capper in this regime (see Fig.~\ref{fig:force_capper_activity}).
\begin{figure*}[t]
    \includegraphics[width=\textwidth]{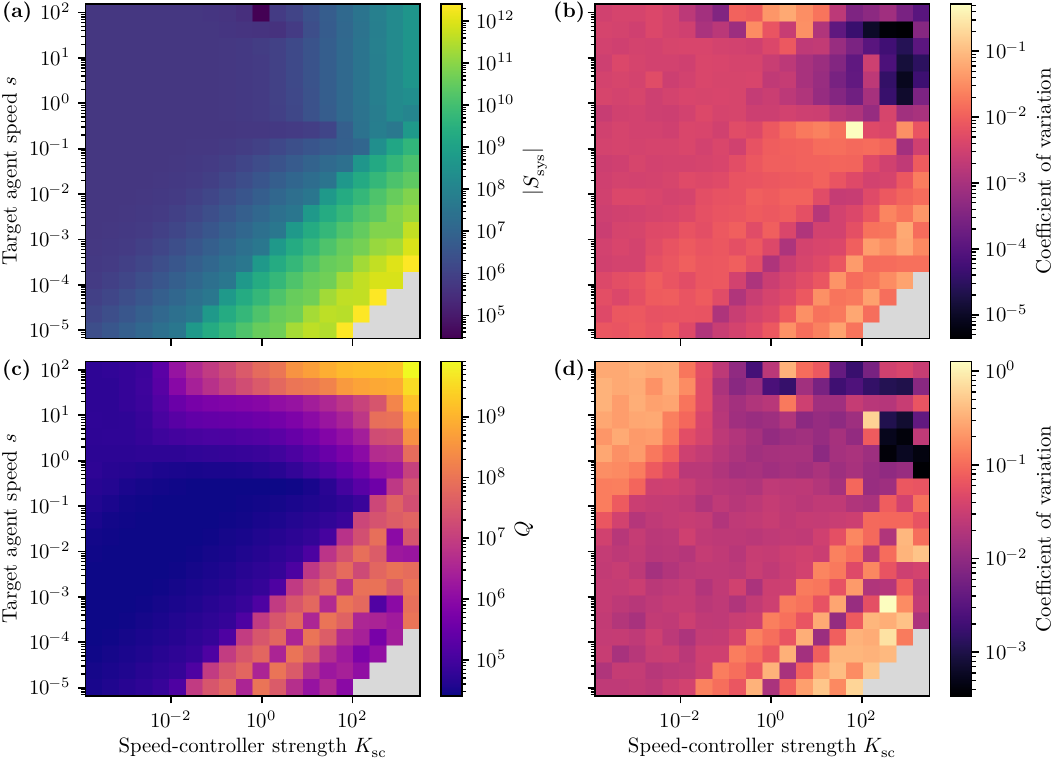}
    \caption{System entropy and transferred heat for the undriven system across the parameter scan of the speed controller force (Eq.~\eqref{eq:speed-controller_force_def}) and the corresponding coefficient of variation. Gray areas indicate crashed simulations due to extreme parameter configurations. The system entropy is negative for all parameter configurations except one, hence the color bar indicates the absolute value of the system entropy.}
    \label{fig:undriven_entropies}
\end{figure*}

In the driven case, the system entropy rate is also negative for all three regimes, but it is larger in the  near-critically damped regime compared to the undriven case, while it is on a similar scale in the underdamped and overdamped regimes. The heat flow is positive for all three regimes, and larger across all three regimes compared to the undriven case, especially in the near-critically damped regime, where it is largest. The dynamics of the system entropy rate and the heat flow in the driven case are again similar in the respective regimes. While the underdamped regime still shows an oscillatory behavior, the near-critically damped regime now also shows a more oscillatory behavior, but it shows pronounced peaks and dips that are largely absent (or only faintly implied) in the underdamped case, suggesting improved temporal resolution. Thus, the near-critically damped regime is not merely intermediate between under- and overdamping; it is the regime in which the driver produces the most temporally structured entropy response. The overdamped regime still shows a rather constant \ac{ep} rate with comparatively large standard deviation, but the dynamics are more noisy compared to the undriven case.

\subsection{Undriven system}

To establish a baseline for the \ac{ep} rates, we first look at a parameter scan of the speed controller force without driving the system.
To compare the \ac{ep} rates across different parameter configurations, we look at the system entropy and heat transferred to the environment,
\begin{equation}
    S_\text{sys} = \int \dot{S}_\text{sys}(t) \text{d}t \quad \text{and} \quad Q = \int \dot{Q}(t) \text{d}t\,,
\end{equation}
as well as the coefficient of variation of these quantities
\begin{equation}
    \text{CV} = \frac{\sigma}{|\mu|}\,,
\end{equation}
where $\sigma$ and $\mu$ are the standard deviation and mean of the respective quantity for a given parameter configuration across different initial conditions.
\begin{figure*}[t]
    \includegraphics[width=\textwidth]{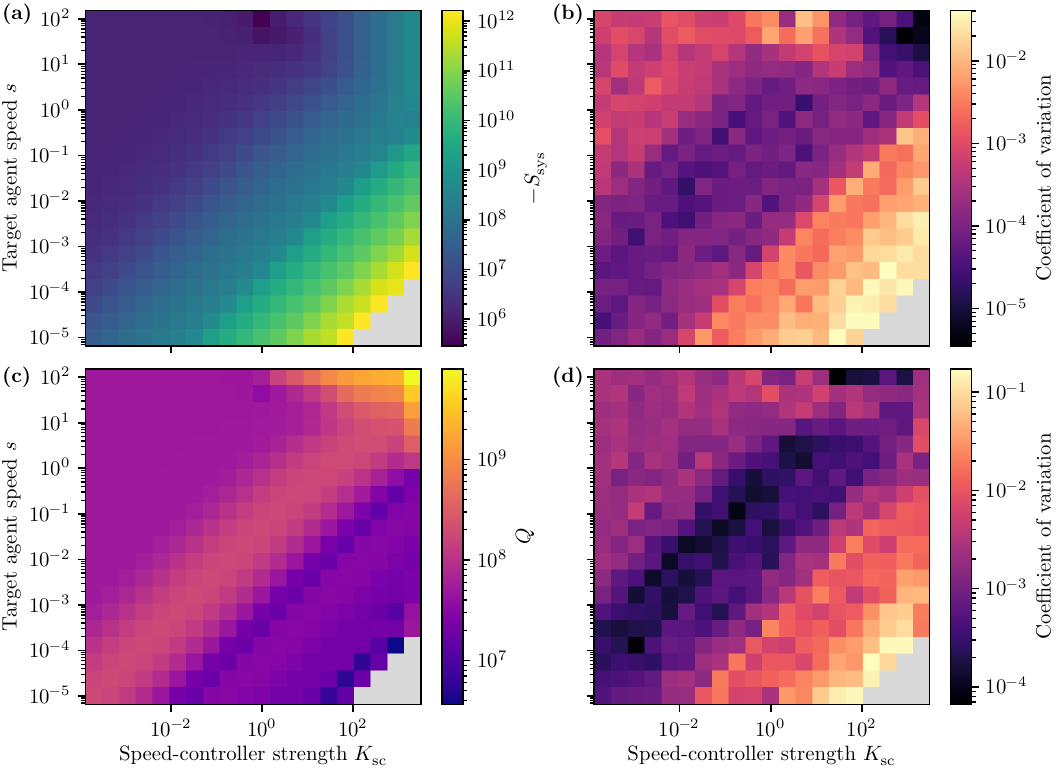}
    \caption{System entropy and transferred heat for the driven system across the parameter scan of the speed controller force (Eq.~\eqref{eq:speed-controller_force_def}) and the corresponding coefficient of variation. Gray areas indicate crashed simulations due to extreme parameter configurations. The system entropy is negative for all parameter configurations, hence the color bar indicates the negative of the system entropy.}
    \label{fig:driven_entropies}
\end{figure*}
The coefficient of variation allows for a better comparison of metrics with different orders of magnitude, which is the case for the system entropy and transferred heat across the parameter scan.
Figure~\ref{fig:undriven_entropies} shows the system entropy and the transferred heat for the undriven system, as well as the corresponding coefficient of variation. The system entropy is negative for all parameter configurations except one ($K_\text{sc} = 856.27$, $s = 0.2637$), hence we use the absolute value to plot it. The positive system entropy can be attributed to a low agent speed compared to the target speed $s$.
The absolute system entropy increases with increasing force strength $K_\text{sc}$ and decreasing target speed $s$, hence it is largest in the overdamped regime and smallest in the underdamped regime.
The transferred heat is positive for all parameter configurations. It has a minimum in the near-critically damped regime, only slightly increasing towards the underdamped regime, but it increases significantly towards the overdamped regime.
Additionally, the coefficient of variation of the transferred heat is small in the near-critically damped regime ($\mathcal{O}(10^{-2})$) and increases towards the underdamped and overdamped regime ($\mathcal{O}(10^{-1})$), which suggests that the transferred heat is more sensitive to the initial conditions in these regimes. Such a clear trend in the coefficient of variation is not visible for the system entropy, however, it is of the same order of magnitude across the parameter scan with a slight increase in the near-critically damped regime ($\mathcal{O}(10^{-2})$), except for some extrema perturbing the color scale.

\subsection{Driven system}

In the driven system, we first look at the same metrics as in the undriven system.
\begin{figure*}[t]
    \includegraphics[width=\textwidth]{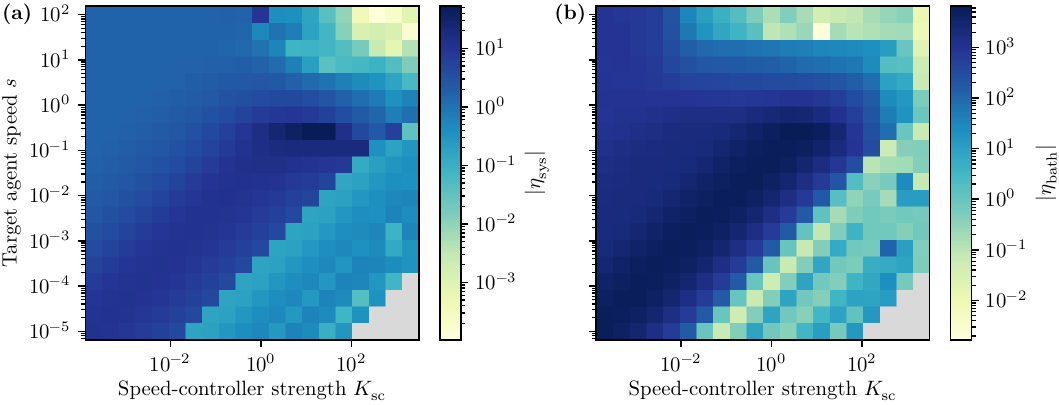}
    \caption{Relative difference between the driven and undriven case for the \textbf{(a)} system entropy and \textbf{(b)} transferred heat across the parameter scan of the speed controller force (Eq.~\eqref{eq:speed-controller_force_def}). Gray areas indicate crashed simulations due to extreme parameter configurations.}
    \label{fig:relative_difference_entropies}
\end{figure*}
Figure~\ref{fig:driven_entropies} shows the system entropy and the transferred heat across the parameter scan for the system driven by a particle that follows a Lorenz-63 trajectory, as well as the corresponding coefficient of variation. The system entropy is negative for all parameter configurations, hence the color bar indicates the negative of the system entropy. Similar to the undriven case, the absolute system entropy increases with increasing force strength $K_\text{sc}$ and decreasing target speed $s$, hence it is largest in the overdamped regime and smallest in the underdamped regime. The absolute system entropy is larger across the parameter scan compared to the undriven case, except for some unphysical configurations in the overdamped regime and in the top right corner.

The transferred heat is positive for all parameter configurations. It has a maximum in the near-critically damped regime. It decreases slowly towards the underdamped regime, but it decreases significantly towards the overdamped regime. Compared to the undriven case, the transferred heat is larger across the parameter scan, except for the overdamped regime, consistent with previous observations that the driver barely influences the dynamics in the overdamped regime, as shown in previous work~\cite{gaimann2025robustly}. Especially the change from a minimum in the near-critically damped regime to a maximum compared to the undriven case is interesting, which suggests that the driver has a significant influence on the transferred heat in this regime.
The coefficient of variation of both the system entropy and the transferred heat is especially small in the near-critically damped regime ($\mathcal{O}(10^{-4})$), which suggests that these metrics are very robust to the initial conditions in this regime. The coefficient of variation increases slightly towards the underdamped and more significantly towards the overdamped regime. However, in the underdamped regime, the coefficient of variation is still quite small ($\mathcal{O}(10^{-3})$). Due to the activity of the force capper in the overdamped regime (see Fig.~\ref{fig:force_capper_activity} (a) and (b)), the results in this regime should be interpreted with caution, which could also explain the larger coefficient of variation.

To visualize the influence of the driver on the system entropy and the transferred heat, we look at the relative difference between the driven and undriven case
\begin{equation}
    \eta_\text{sys} = \frac{S_\text{sys}^\text{driven} - S_\text{sys}^\text{undriven}}{S_\text{sys}^\text{undriven}}
\end{equation}
and
\begin{equation}
    \eta_\text{bath} = \frac{Q^\text{driven} - Q^\text{undriven}}{Q^\text{undriven}}
\end{equation}
across the parameter scan, which is shown in Fig.~\ref{fig:relative_difference_entropies}.
Since the magnitudes of the undriven entropy and transferred heat values are greater than $10^{4}$, the relative difference is well-defined across the parameter scan and is not dominated by a small denominator.
The relative difference of the system entropy peaks in the near-critically damped regime with a maximum of $40$. The relative difference of the transferred heat also peaks in the near-critically damped regime, but reaches about $6 \cdot 10^{3}$, indicating that the driver influences the transferred heat much more strongly than the system entropy. Since the transferred heat quantifies dissipation, this pattern indicates stronger driver induced dissipation in the near-critically damped regime. Interpreted in terms of information balance, it is consistent with higher driver information injection in this regime and a larger fraction of that information being lost to dissipation. A plausible implication is that increased driver induced dissipation accompanies the improved reservoir performance reported for this regime~\cite{gaimann2025robustly}, although this link remains interpretive. The undriven transferred heat landscape shows that the near-critically damped regime is not simply the most dissipative regime intrinsically (see Fig.~\ref{fig:undriven_entropies}(c)). Its computational relevance emerges only once the system is driven.
\begin{figure*}[t]
    \includegraphics[width=\textwidth]{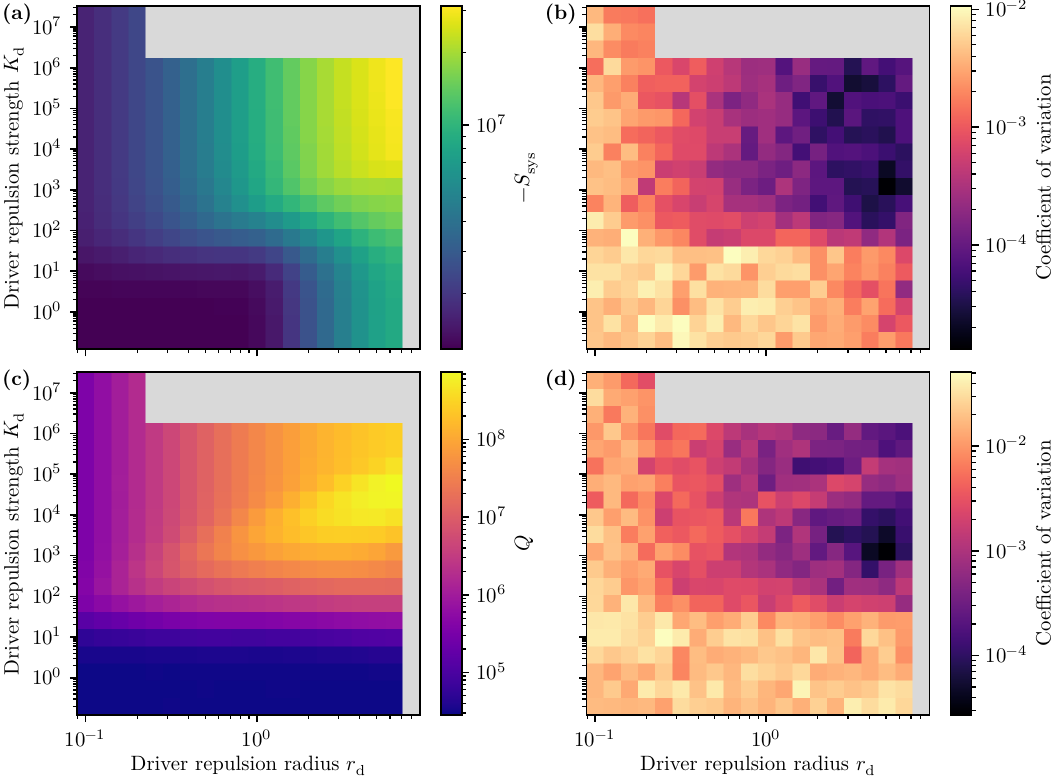}
    \caption{System entropy and transferred heat for the driven system across the driver force scan (Eq.~\eqref{eq:driver_force}) and the corresponding coefficient of variation. Gray areas indicate crashed simulations due to extreme parameter configurations. The system entropy is negative for all parameter configurations, hence the color bar indicates the negative of the system entropy.}
    \label{fig:entropies_driver_repulsion}
\end{figure*}

For a broader analysis of the entropy metrics, we performed a second parameter scan across the driver force strength $K_\text{d}$ and repulsion radius $r_\text{d}$, with $K_\text{sc} = 0.4138$ and $s = 0.0483$ (near-critically damped regime), which is shown in Fig.~\ref{fig:entropies_driver_repulsion}.
The system entropy is negative for all parameter configurations, hence the color bar indicates the negative of the system entropy. The absolute system entropy increases with increasing driver force strength $K_\text{d}$ and increasing repulsion radius $r_\text{d}$, hence it is largest in the top right corner and smallest in the bottom left corner. The transferred heat is positive for all parameter configurations. It also increases with increasing driver force strength $K_\text{d}$ and increasing repulsion radius $r_\text{d}$. The coefficient of variation of both the system entropy and the transferred heat is small across the parameter scan ($\mathcal{O}(10^{-2})$), and decreases with increasing driver force strength $K_\text{d}$ and increasing repulsion radius $r_\text{d}$, which suggests that the system is more robust to the initial conditions.

\subsection{Driver work}

Since the driver seems to have a significant influence on the \ac{ep} rate, especially in the near-critically damped regime, we want to further quantify this influence by calculating the work of the driver on the system.
\begin{figure*}[t]
    \includegraphics[width=\textwidth]{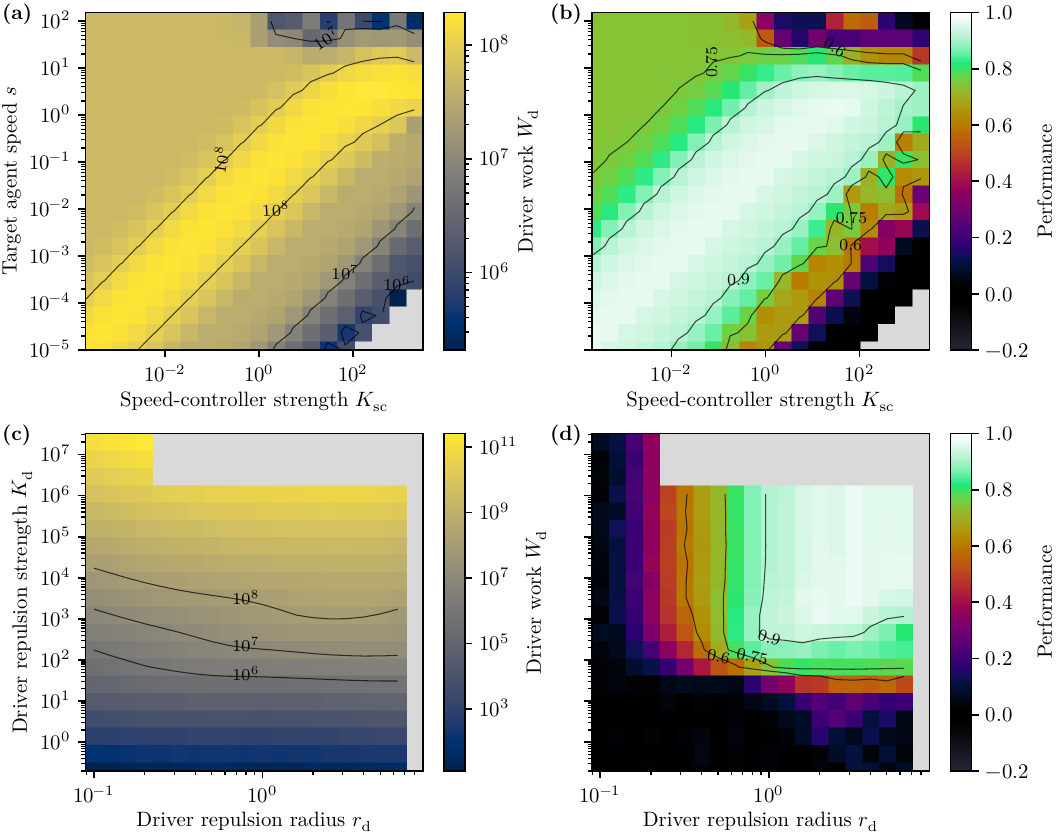}
    \caption{Work of the driver on the system and performance across both parameter scans. Gray areas indicate crashed simulations due to extreme parameter configurations. The work is shown in log scale to better visualize the differences across the parameter scan. The performance is measured via the Pearson correlation coefficient between the system prediction and the target signal, which is the $x$-coordinate of the Lorenz-63 trajectory.}
    \label{fig:work_and_performance_driven_and_repulsion}
\end{figure*}
The work of the driver is defined in Eq.~\eqref{eq:driver-work} and can be computed for each particle $k$ via
\begin{equation}
    W_{\text{d}, k} = \int \mathbf{F}_{\text{d}, k}(t) \D \mathbf{s} = \int \mathbf{F}_{\text{d}, k}(t) \cdot \mathbf{v}_k(t) \D t\,,
\end{equation}
where $\mathbf{F}_{\text{d}, k}(t)$ is the driver force acting on particle $k$ at time $t$ and $\mathbf{v}_k(t)$ is the corresponding velocity of particle $k$. The total work of the driver on the system is then calculated by the sum of the work of the driver on all particles,
\begin{equation}
    W_\text{d} = \sum_k W_{\text{d}, k}\,.
\end{equation}
This work is shown across both parameter scans in Fig.~\ref{fig:work_and_performance_driven_and_repulsion} (a) and (c).
In panels (b) and (d) the performance of the system as a reservoir computer is shown, measured via the Pearson correlation coefficient~\cite{pearson1896mathematical} between the system prediction and the target signal, which is the $x$-coordinate of the Lorenz-63 trajectory (compare and see Ref.~\cite{gaimann2025robustly,gaimann2026optimal} for details). We want to highlight, that we reach higher performances (maximum of $0.97$) across the parameter scans compared to previous work~\cite{gaimann2025robustly,gaimann2026optimal}. Even though the smaller time step size $\text{d}t=0.002$ was already investigated in Ref.~\cite{gaimann2025robustly}, we use a driving signal generated according to Appendix A of Ref.~\cite{gaimann2025robustly} with a specified mean speed, which is different from the driving signal used there. Additionally, we only record every tenth time step, resulting in fewer data for the reservoir training. We suspect that the smaller time step size and the different driving signal, which matches the timescales of the system better, result in richer observed reservoir states. Hence, there might be a more consistent input-output mapping, which could explain the improved performance. However, a more detailed investigation of the influence of the time step size and the driving signal on the reservoir performance is beyond the scope of this work.

For the speed-controller force scan, the driver work closely resembles the performance landscape, with a maximum in the near-critically damped regime, a gradual decrease toward the underdamped regime, and a stronger decrease toward the overdamped regime. This suggests that work input by the driver is an important factor for reservoir performance. However, in the driver-force scan, the work landscape does not resemble the performance landscape as closely, indicating that work input alone is insufficient to characterize computational performance.
It is also important how the driver force is distributed across the system. Therefore, we calculate the driver force concentration
\begin{equation}
    c_{\text{d}} = \frac{1}{N_t} \sum_{n=1}^{N_t} \frac{\max_k ||\mathbf{F}_{\text{d}, k}(t_n)||}{\sum_k ||\mathbf{F}_{\text{d}, k}(t_n)||}\,,
    \label{eq:driver-force-concentration}
\end{equation}
which is close to 1 if the driver force is concentrated on a single particle and close to $1/N$ if the driver force is evenly distributed across all $N$ particles. Time steps for which the total driver force vanishes, which is rarely the case, are excluded from the average.
\begin{figure}
    \includegraphics[width=\columnwidth]{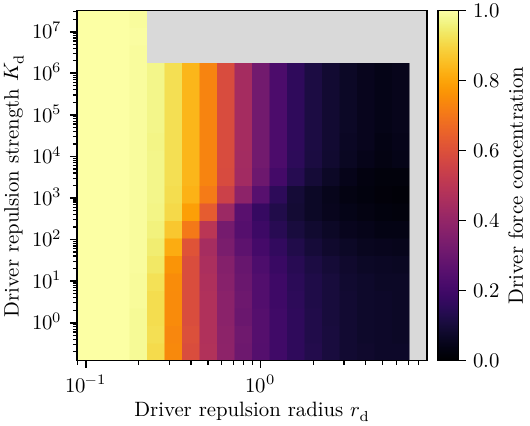}
    \caption{Driver force concentration (Eq.~\eqref{eq:driver-force-concentration}) across the driver force scan (Eq.~\eqref{eq:driver_force}). Gray areas indicate crashed simulations due to extreme parameter configurations. The driver force concentration is close to 1 if the driver force is concentrated on a single particle and close to $1/N$ if the driver force is evenly distributed across all $N$ particles. For larger radius $r_\text{d}$, the driver force is more evenly distributed across the system, while for smaller radius $r_\text{d}$, the driver force is concentrated on a single particle.}
    \label{fig:force_concentration_driver_repulsion}
\end{figure}
The driver force concentration across the driver force scan is shown in Fig.~\ref{fig:force_concentration_driver_repulsion} and shows that for larger radius $r_\text{d}$, the driver force is more evenly distributed across the system, while for smaller radius $r_\text{d}$, it is concentrated on a single particle. Combining work input with this concentration measure gives the heuristic distributed-work metric $W_\text{d} \cdot (1 - c_\text{d})$. This metric is large when the driver performs substantial work and this work is not concentrated on a single particle. Compared with $W_\mathrm{d}$ alone, $W_\mathrm{d}(1-c_\mathrm{d})$ more closely resembles the performance landscape in Fig.~\ref{fig:work_and_performance_driven_and_repulsion} (d).

To further quantify the relationship between the thermodynamic metrics and the performance, we calculate the Pearson correlation coefficient~\cite{pearson1896mathematical} and the Spearman rank correlation coefficient~\cite{spearman1904proof} between the performance and selected metrics across both parameter scans, which are shown in Tab.~\ref{tab:performance_metric_correlations_speed_controller} and~\ref{tab:performance_metric_correlations_repulsion}.
\begin{table}
\caption{Correlations between performance and selected metrics for the driven system across the speed controller force scan (Eq.~\eqref{eq:speed-controller_force_def}). Only parameter configurations, where the force capper acts on $\langle N_\text{cap} \rangle \leq 60$ particles on average per time step, are considered. Hence, the overdamped regime is excluded from the correlation analysis to keep the influence of the force capper on the results small.}
\centering
\begin{tabular}{l|rr}
Metric & Pearson $r$ & Spearman $\rho$ \\
\hline
$S_\mathrm{sys}$ & -0.1319 & -0.5454 \\
$Q$ & -0.4625 & 0.5051 \\
$W_\mathrm{d}$ & 0.6666 & 0.8693 \\
$W_\mathrm{d} \cdot (1 - c_\mathrm{d})$ & 0.6612 & 0.9023 \\
CV($S_\mathrm{sys}$) & -0.5252 & -0.6145 \\
CV($Q$) & -0.4512 & -0.7161 \\
$\eta_\mathrm{sys}$ & 0.4426 & 0.8085 \\
$\eta_\mathrm{bath}$ & 0.5201 & 0.7547 \\
$S_\mathrm{sys}^\mathrm{undriven}$ & 0.1869 & -0.3922 \\
$Q^\mathrm{undriven}$ & -0.5991 & -0.5391 \\
\end{tabular}
\label{tab:performance_metric_correlations_speed_controller}
\end{table}
\begin{table}
\caption{Correlations between performance and selected metrics for the driven system across the driver force scan (Eq.~\eqref{eq:driver_force}). Only parameter configurations, where the force capper acts on $\langle N_\text{cap} \rangle \leq 60$ particles on average per time step, are considered.}
\centering
\begin{tabular}{l|rr}
Metric & Pearson $r$ & Spearman $\rho$ \\
\hline
$S_\mathrm{sys}$ & -0.7438 & -0.9098 \\
$Q$ & 0.5446 & 0.9513 \\
$W_\mathrm{d}$ & 0.0301 & 0.7001 \\
$W_\mathrm{d} \cdot (1 - c_\mathrm{d})$ & 0.3449 & 0.8938 \\
CV($S_\mathrm{sys}$) & -0.7930 & -0.9182 \\
CV($Q$) & -0.8187 & -0.9023 \\
\end{tabular}
\label{tab:performance_metric_correlations_repulsion}
\end{table}
The Pearson correlation coefficient measures the linear correlation between two variables, while the Spearman rank correlation coefficient measures the monotonic relationship between two variables. To reduce the influence of the force capper on the correlation analysis, we only consider parameter configurations, where the force capper acts on $\langle N_\text{cap} \rangle \leq 60$ particles on average per time step. Hence, the overdamped regime is excluded from the correlation analysis in the speed controller force scan.

Across the speed-controller scan, raw system entropy and transferred heat show only moderate monotonic correlations with performance. In contrast, driver work and driven/undriven entropy ratios show consistently stronger correlations (see Tab.~\ref{tab:performance_metric_correlations_speed_controller}). The coefficients of variation of both system entropy and transferred heat correlate negatively with performance, supporting the interpretation that a more robust system, i.e., one that is less sensitive to initial conditions, performs better as a reservoir computer. Notably, the heuristic distributed-work metric and the driver work alone yield the highest Spearman correlation, closely followed by the relative difference in system entropy. We observed, that the Pearson correlation increases substantially, when considering the logarithmic ratio of the driven and undriven quantities. Since the Pearson correlation is comparatively low for the other metrics, the relationship with performance appears to be monotonic rather than linear, which is captured by the Spearman coefficients. The Spearman correlation of the undriven transferred heat is surprisingly high (anti-correlated with performance), since the undriven transferred heat is not influenced by the driver and thus should not be related to the performance.

For the driver-force scan, the undriven system entropy and the transferred heat are constant across the parameter space, since the driver does not affect the dynamics in the undriven case; consequently, a relative difference would amount only to a rescaling and would not change the correlation. In this scan, the system entropy correlates negatively with the performance, and all metrics correlate well with the performance regarding the Spearman correlation, only the driver work is comparatively lower (see Tab.~\ref{tab:performance_metric_correlations_repulsion}). However, if we consider the heuristic distributed-work metric, the correlation is considerably higher. Again, the coefficient of variation of both system entropy and transferred heat correlate negatively with the performance. Overall, the system entropy and transferred heat serve as a strong empirical predictor for the performance of the system as a reservoir computer. The higher correlations in the driver force scan compared to the speed controller force scan could be due to the unphysical behavior in the top right corner in the speed controller force scan.

The same correlation analysis was repeated for four additional attractor-based driver signals, yielding the same qualitative trends (Appendix~\ref{app:additional-drivers}).


\section{Discussion and Outlook}
\label{sec:discussion}

A central question in physical reservoir computing is whether good computational regimes can be identified from physical observables rather than from task performance alone. The entropy-production analysis presented here suggests that this is possible, but only in a qualified sense. In the active-matter reservoir studied here, computational performance is not controlled by the absolute magnitude of entropy production. Rather, the relevant signature is the driver-induced change in the thermodynamic response. Entropy production is therefore not a universal scalar proxy for performance; its diagnostic value comes from separating intrinsic phase-space contraction, bath dissipation, and driver-induced changes.

This distinction is visible in the contrast between the undriven and driven systems. In the undriven system, the near-critically damped band is not the most dissipative regime; transferred heat is instead minimal there. Under driving, however, this same region develops the strongest increase in transferred heat and work input. This contrast indicates that near-critical damping is not distinguished by high intrinsic dissipation, but by its sensitivity to external perturbations.
Quantitatively, the driven near-critically damped regime shows the largest driver-induced dissipation (Fig.~\ref{fig:relative_difference_entropies}), whereas in the undriven scan the transferred heat has a minimum in the near-critically damped band (Fig.~\ref{fig:undriven_entropies}). The time-resolved entropy rates provide the dynamical counterpart to this observation. The underdamped regime remains more oscillatory, while the overdamped regime converts the drive less effectively into useful reservoir-state variation. Consistent with this, the driven near-critically damped time series shows pronounced peaks and dips (Fig.~\ref{fig:entropy_time_series_driven}), indicating a more temporally structured response to the driver rather than merely larger magnitudes. The near-critically damped regime is most favorable for the prediction task: it shows the strongest driver-induced thermodynamic response and dissipation, while the underdamped regime remains more oscillatory and the overdamped regime converts the drive less effectively into useful reservoir-state variation. A plausible interpretation is that the near-critically damped regime permits driver perturbations to be converted into collective swarm responses, whereas the underdamped regime stores more of the injected energy in oscillatory motion and the overdamped regime suppresses the dynamical response.

The swarm model studied here is established in the reservoir-computing literature~\cite{lymburnReservoirComputingSwarms2021a,gaimann2025robustly,gaimann2026optimal}; the main contribution of this work is the thermodynamic analysis of its entropy production. Separating system entropy rate and heat flow clarifies how phase-space contraction and dissipation vary across underdamped, near-critically damped, and overdamped regimes. The system entropy estimator derived from the generalized Liouville equation provides a measure of phase-space contraction. Because it relies only on the deterministic flow field, it applies to arbitrary systems of first-order \acp{ode} and is not tied to the specific swarm model investigated here.
In contrast, the heat flow depends on the identification of the Hamiltonian and non-Hamiltonian force split, and is therefore more model-specific. Together, these quantities provide complementary perspectives: a broadly applicable measure of phase-space contraction and a physically motivated measure of dissipation when the required force decomposition is meaningful.

The thermodynamic view also helps evaluate reservoir computational abilities and their robustness. Driver work tracks performance in the speed-controller scan, but not universally. Entropy ratios and driven-undriven contrasts are more consistent across scans, suggesting that performance depends on the driver-induced thermodynamic response rather than absolute entropy production (Tabs.~\ref{tab:performance_metric_correlations_speed_controller} and~\ref{tab:performance_metric_correlations_repulsion}). The same qualitative trends are obtained for four additional attractor-based driver signals (Appendix~\ref{app:additional-drivers}), suggesting that the observed relationship between driver-induced thermodynamic response and performance is not specific to the Lorenz-63 input. In the speed-controller scan, the driver work closely tracks the performance landscape, with strong correlations (Tab.~\ref{tab:performance_metric_correlations_speed_controller}), indicating that work input is a key ingredient for good computation, whereas the driver-force scan indicates that work alone is insufficient and that it is also relevant how distributed the driver force is across the swarm. Additionally, the driven-undriven entropy differences also align consistently with the performance.
The stronger Spearman than Pearson correlations indicate a largely monotonic relationship rather than a single linear scaling.

The robustness analysis supports the same interpretation. The coefficient of variation of both system entropy and transferred heat correlates negatively with performance and is smallest in the near-critically damped regime. Thus, the regimes with the strongest computational performance are also those in which the thermodynamic response is most robust to initial conditions. This reinforces the view that entropy production is informative not only through its average magnitude, but also through the reproducibility of the driver-induced response.

These interpretations come with limitations. The transferred heat relies on a model-specific Hamiltonian and non-Hamiltonian force split and a passive-bath assumption, so alternative coupling choices could change its absolute magnitude. We further confirmed that Eq.~\ref{eq:eprate_total} is positive only where the driver dominates the system's dissipative response, turning negative when driving is weak or absent (small $K_\text{d}$, small $r_\text{d}$, undriven case, overdamped regime). This is consistent with the neglected constituent entropy $\dot{S}_\text{cons}$ becoming non-negligible precisely where the driver's contribution to dissipation is small. The force capper affects overdamped configurations and modifies the intended force decomposition, so transferred heat calculations in this region should be interpreted cautiously. Finally, correlations between performance and entropy or work do not establish causality. They indicate predictive associations rather than mechanistic proof.

Future work should examine thermodynamically consistent bath couplings to enable a full entropy balance and sharpen the interpretation of dissipation, and alternative time-reversal prescriptions to connect this deterministic setting more directly to stochastic-thermodynamic notions of irreversibility. It would also be useful to relate entropy production to information-theoretic measures such as memory or information transfer, and to test whether entropy and dissipation can serve as a task-agnostic proxy for computational ability across other reservoir classes.


\begin{acknowledgments}

We thank Max Weinmann and Lasse Schulz for helpful and stimulating discussions.
Funded by Deutsche Forschungsgemeinschaft (DFG, German Research Foundation) under Germany's Excellence Strategy -- EXC 2075 -- 390740016 and by the Ministry of Science, Research and the Arts Baden-W\"{u}rttemberg (Az.~33-7533-9-19/54/5) in \enquote{K\"{u}nstliche Intelligenz \& Gesellschaft: Reflecting Intelligent Systems for Diversity, Demography and Democracy} (IRIS3D). We acknowledge the support of the Stuttgart Center for Simulation Science (SimTech), the Interchange Forum for Reflecting on Intelligent Systems (IRIS) at the University of Stuttgart, the International Max Planck Research School for Intelligent Systems (IMPRS-IS), and the Heidelberg Academy of Sciences and Humanities.

\end{acknowledgments}


\section*{Author contributions}

P.E., A.K.~and M.K.~conceived the idea, P.E.~and M.K.~directed the project. P.E.~and H.K.~designed the study, performed the simulations and analyzed the data. M.U.G.~provided the software framework and software support. P.E.~and A.K.~wrote the paper. P.E.~and M.K.~revised the paper.  M.K.~is the principal investigator of the project. All the authors discussed the results and commented on the manuscript.


\section*{Data availability statement}

The data that support the findings of this study are openly available at~\cite{DARUS-6343_2026}.

\bibliography{references}

\appendix

\section{Influence of the Force Capper}
\label{app:force-capper}

To verify that the force capper does not significantly influence the results, we look at the activity of the force capper across the parameter scans.
\begin{figure}
    \includegraphics[width=\columnwidth]{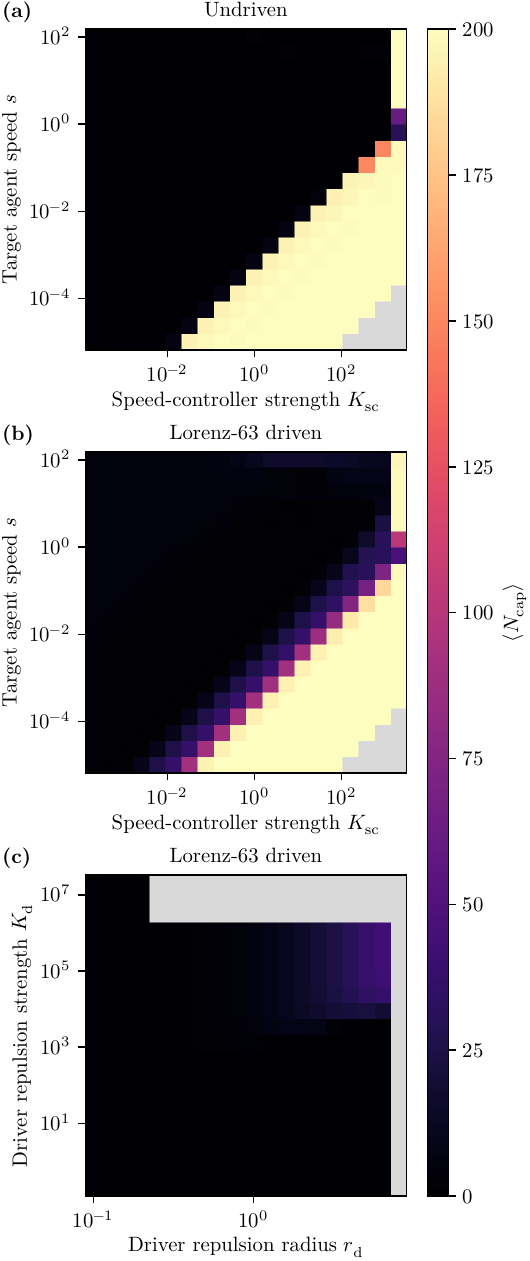}
    \caption{Average number of particles per time step on which the force capper is active across the parameter scans. Gray areas indicate crashed simulations. The panels \textbf{(a)} and \textbf{(b)} show the force capper activity across the speed controller force scan (Eq.~\ref{eq:speed-controller_force_def}) for the undriven and driven case, respectively, and \textbf{(c)} across the driver force scan (Eq.~\ref{eq:driver_force}).}
    \label{fig:force_capper_activity}
\end{figure}
Therefore, we define the average number of particles per time step on which the force capper is active via
\begin{equation}
    \langle N_\mathrm{cap} \rangle = \frac{1}{N_t} \sum_i^{N_t} N_\mathrm{cap}(t_i)\,,
\end{equation}
where $N_\mathrm{cap}(t_i)$ is the number of particles on which the force capper is active at time step $t_i$.
Figure~\ref{fig:force_capper_activity} shows that the force capper is only active on a very small number of particles in the speed controller force scans in the underdamped and near-critically damped regimes, which suggests that it does not significantly influence the observed dynamics and hence the calculated \ac{ep} rate in these regimes. In the overdamped regime, the force capper is active on a larger number of particles, which can be attributed to the large force strength $K_\text{sc}$ in this regime, as well as the fact that the particles are moving more slowly and hence are more likely to get close to each other.
In the driver force scan, the force capper is only active on a few particles for large repulsion radii $r_\text{d}$ and large force strengths $K_\text{d}$. For the non-driven case in the underdamped and near-critically damped regimes and for the driver force scan outside the largest $K_\text{d}$-$r_\text{d}$ corner, the force capper does not activate at all.


\section{Robustness across additional driver signals}
\label{app:additional-drivers}

In the main text, reservoir performance and thermodynamic response are analyzed using the Lorenz-63 system as the external driver. To test whether the observed relationship between entropy production, driver work, and reservoir performance is specific to this choice of input signal, we repeated the analysis for four additional attractor-based driver signals, namely the Chua, R\"{o}ssler, H\'{e}non-Heiles, and Lorenz-96 systems. The same two parameter scans were performed for each driver: the speed-controller scan and the driver-force scan. For each scan, we computed the same thermodynamic and performance metrics as in the main text and evaluated their Pearson and Spearman correlations with prediction performance.
\begin{table*}
\centering
\caption{Pearson and Spearman correlations between performance and selected metrics for the speed controller force scan (Eq.~\eqref{eq:speed-controller_force_def}). Only parameter configurations, where the force capper acts on $\langle N_\text{cap} \rangle \leq 60$ particles on average per time step, are considered. Hence, the overdamped regime is excluded from the correlation analysis to keep the influence of the force capper on the results small.}
\begin{tabular}{l|rr|rr|rr|rr}
Metric & \multicolumn{2}{c}{Chua} & \multicolumn{2}{c}{R\"{o}ssler} & \multicolumn{2}{c}{H\'{e}non-Heiles} & \multicolumn{2}{c}{Lorenz-96} \\
 & Pearson $r$ & Spearman $\rho$ & Pearson $r$ & Spearman $\rho$ & Pearson $r$ & Spearman $\rho$ & Pearson $r$ & Spearman $\rho$ \\
\hline
$S_\mathrm{sys}$ & -0.0184 & -0.3928 & -0.0029 & -0.4898 & 0.2899 & -0.2168 & -0.2005 & -0.5800 \\
$Q$ & -0.6180 & 0.4041 & -0.6160 & 0.4967 & -0.5375 & 0.5074 & -0.2999 & 0.5290 \\
$W_\mathrm{d}$ & 0.3928 & 0.7505 & 0.3516 & 0.7887 & 0.3288 & 0.6927 & 0.7869 & 0.8345 \\
$W_\mathrm{d} \cdot (1 - c_\mathrm{d})$ & 0.3711 & 0.7469 & 0.3449 & 0.7858 & 0.3143 & 0.6897 & 0.7772 & 0.8383 \\
CV($S_\mathrm{sys}$) & -0.5989 & -0.6908 & -0.2786 & -0.5898 & -0.4578 & -0.6888 & -0.5204 & -0.6690 \\
CV($Q$) & -0.3763 & -0.7139 & -0.2269 & -0.6871 & -0.3205 & -0.7892 & -0.6252 & -0.7382 \\
$\eta_\mathrm{sys}$ & 0.2844 & 0.7679 & 0.2815 & 0.7772 & 0.2597 & 0.7399 & 0.5408 & 0.7881 \\
$\eta_\mathrm{bath}$ & 0.3132 & 0.7747 & 0.3134 & 0.8465 & 0.2972 & 0.8093 & 0.6365 & 0.7636 \\
$S_\mathrm{sys}^\mathrm{undriven}$ & 0.2150 & -0.2488 & 0.3542 & -0.3199 & 0.3019 & -0.2136 & 0.1209 & -0.4219 \\
$Q^\mathrm{undriven}$ & -0.7255 & -0.6602 & -0.7058 & -0.6003 & -0.7534 & -0.6671 & -0.5232 & -0.5269 \\
\end{tabular}
\label{tab:additional_driver_correlations_speed_controller}
\end{table*}
\begin{table*}
\centering
\caption{Pearson and Spearman correlations between performance and selected metrics for the driver force scan (Eq.~\eqref{eq:driver_force}). Only parameter configurations, where the force capper acts on $\langle N_\text{cap} \rangle \leq 60$ particles on average per time step, are considered.}
\begin{tabular}{l|rr|rr|rr|rr}
Metric & \multicolumn{2}{c}{Chua} & \multicolumn{2}{c}{R\"{o}ssler} & \multicolumn{2}{c}{H\'{e}non-Heiles} & \multicolumn{2}{c}{Lorenz-96} \\
 & Pearson $r$ & Spearman $\rho$ & Pearson $r$ & Spearman $\rho$ & Pearson $r$ & Spearman $\rho$ & Pearson $r$ & Spearman $\rho$ \\
\hline
$S_\mathrm{sys}$ & -0.6709 & -0.9329 & -0.5273 & -0.8986 & -0.5804 & -0.9124 & -0.7562 & -0.9503 \\
$Q$ & 0.4957 & 0.9087 & 0.3919 & 0.9717 & 0.4448 & 0.9642 & 0.5649 & 0.9320 \\
$W_\mathrm{d}$ & -0.0049 & 0.6085 & 0.1736 & 0.7201 & 0.0710 & 0.7124 & 0.0946 & 0.6605 \\
$W_\mathrm{d} \cdot (1 - c_\mathrm{d})$ & 0.2976 & 0.8536 & 0.2267 & 0.8936 & 0.2552 & 0.8927 & 0.3214 & 0.8570 \\
CV($S_\mathrm{sys}$) & -0.8314 & -0.9093 & -0.7258 & -0.8613 & -0.7264 & -0.8520 & -0.7743 & -0.8918 \\
CV($Q$) & -0.7956 & -0.8633 & -0.6525 & -0.8326 & -0.6926 & -0.8528 & -0.7687 & -0.8588 \\
\end{tabular}
\label{tab:additional_driver_correlations_repulsion}
\end{table*}

The additional drivers support the main conclusions. Across the tested attractors, performance is not explained by the absolute magnitude of entropy production alone. Instead, the most consistent associations are obtained for quantities that measure the driver-induced thermodynamic response, such as driven-undriven entropy contrasts and driver work. In the speed-controller scans, driver work remains strongly associated with performance, consistent with the main Lorenz-63 results. In the driver-force scans, work alone is less consistently predictive, while metrics that combine work input with the structure of driver coupling provide a more robust description. Thus, the qualitative relation between thermodynamic response and computational performance is not specific to the Lorenz-63 driver.

Tables~\ref{tab:additional_driver_correlations_speed_controller} and~\ref{tab:additional_driver_correlations_repulsion} summarize the correlation analysis for the additional attractors. The stronger and more consistent Spearman correlations indicate that the relevant relationship is often monotonic rather than described by a single linear scaling. This is consistent with the main results, where entropy production and driver work are useful diagnostic quantities but do not constitute universal scalar proxies for performance.

\end{document}